\documentclass[twocolumn,preprintnumbers,amsmath,amssymb]{revtex4}

\usepackage{graphicx}
\usepackage{bm}

\newcommand{\beq}{\begin{equation}}
\newcommand{\eeq}{\end{equation}}
\newcommand{\bea}{\begin{eqnarray}}
\newcommand{\eea}{\end{eqnarray}}

 \let\b=\beta     
       
                 \let\r=\rho
\let\s=\sigma \let\t=\tau   \let\f=\varphi

\begin{document}

\title
{Phase transitions and topology in $2+k$ $XY$ mean-field models}

\author{L. Angelani$^1$ and G. Ruocco$^{2,3}$}

\affiliation{
$^1$Research center SMC INFM-CNR, c/o Universit\`a
di Roma ``La Sapienza,'' I-00185, Roma, Italy \\
$^2$Research center Soft INFM-CNR, c/o Universit\`a di
Roma ``La Sapienza,'' I-00185, Roma, Italy \\
$^3$Dipartimento di Fisica, Universit\`a di Roma ``La Sapienza,''
I-00185, Roma, Italy
}

\begin{abstract}
The thermodynamics and topology of mean-field models 
with $2+k$ body interaction terms (generalizing $XY$ model)
are derived. Focusing on two particular cases ($2+4$ and
$2+6$ body interaction terms), a comparison between thermodynamic
(phase transition energy, thermodynamically forbidden energy regions) and
topological (singularity and curvature of saddle
entropy) properties is performed. We find that 
{\it i}) a topological change is present at the phase transition energy;
however, 
{\it ii}) only one topological
change occurs, also for those models exhibiting two phase
transitions; 
{\it iii}) the order of a phase transition is not completely 
signaled by the curvature of topological quantities.
\end{abstract}

\maketitle

\section{Introduction}
In the recent years different authors suggested a possible {\it
topological} approach to the study of the phase transitions.
Within this approach it has been suggested that any thermodynamic
phase transition mirrors a topological change of the potential 
energy hypersurface \cite{cccp,cpc}, i.e. a change in the
topology of certain submanifolds in configuration space (for a
recent review see Ref.s \cite{kastner_RMP,pettini_book}). 
More specifically, the
energy density $e$ where a thermodynamic phase transition takes
place ($e=e_c$ in the microcanonical, or $e=e(T_c)$ in the
canonical, ensemble) has been conjectured to be the same where a
topological change of the submanifold $M_e = \{q|V(q) \leq N e\}$
appears (here $q$ are the generalized coordinates, $V(q)$ the
potential energy function, $N$ the number of degrees of freedom).
This ``topological hypothesis'' has been subsequently formalized
in a theorem which, however, applies only to a strict class of
systems, described by smooth, bounded below, confining, finite
range potentials \cite{fra_pet,FP_theo1,FP_theo2}. 
This theorem states that a
topological change is a necessary condition for the presence of a
phase transition. However the study of model systems fulfilling
the requirements of the theorem is a very hard task, due to
analytical difficulty to solve the thermodynamics and/or topology
or to numerically calculate topological quantities. For this
reason, beside few cases \cite{fra_2000,rrs,XY}, many works have
been devoted to the study of tractable model systems
not-fulfilling the theorem hypotheses
\cite{XY,kastner,gri_mos,teix,ktrig,phi4,1d}.
For such models a variety of results has been obtained, some in
agreement and some not with the topological hypothesis (see Table
I of Ref. \cite{kastner_RMP} for a summary). Among the others, two
particular interesting questions remain to be answered.

The first question concerns the presence of topological changes in
correspondence of phase transition energy values. Is a topological
change in those systems undergoing a phase transition always
present? And more, is the energy $e_\t$ where topological change
are observed coincident to $e_c$, the thermodynamic phase
transition energy? It is well established that for certain models
(not-fulfilling the hypotheses of the theorem) the two energies
are not coincident. Two different mechanisms have been proposed to
be responsible of this discrepancy: a maximization procedure of a
smooth function generating a phase transition 
(as opposite mechanism with respect to the topological one)
\cite{kas1} or an underlying saddle-dominated dynamics for which
the relevant topological energy level is not the instantaneous
potential energy, rather it is the potential energy where are
located the saddles of $V(q)$ mostly visited at the thermodynamic
phase transition state point (``weak'' topological hypothesis)
\cite{phi4,1d}. It is worth to mention that in all the models
analyzed so far there is always a topological change (although at
an energy not coincident with the thermodynamic one) in the
presence of a phase transition.

The second open question regards the possibility to infer the
order of the phase transition from the curvature properties of 
topological invariants such as the Euler characteristic. In other
words, is there a one-to-one correspondence between curvatures of
thermodynamic entropy and some topological invariant quantity? In
previous studies of toy models ($k$-trigonometric model) \cite{ktrig},
capable of switching between first- and second-order phase
transition by tuning a control parameter, a positive answer to
this question was given. Specifically, we observed positive
curvature of saddle entropy for systems undergoing first-order
phase transition, while a standard negative curvature accompanied
the second order transitions. The relevant control parameter of
the $k$-trigonometric model is the number $k$ of interacting bodies in the
Hamiltonian which, in turn, depends on the relative phases of
these $k$-interacting bodies.

In this paper we study a class of mean-field models 
with  different many-body interaction terms, 
which, therefore, can be named as $k+k'$
model. Both thermodynamics and topology are analytically
tractable, so a direct comparison can be made between the two. Two
particular model systems will be analyzed (2+4 and 2+6), which
manifest a rich phase diagram. Our main findings are the
following. {\it i}) There is always a topological change at the
same energy level for all the model systems, corresponding to the
paramagnetic energy at which the phase transition (first or second
order, depending on the model) takes place. {\it ii}) For one of
the two models two phase transitions are observed
(paramagnetic-magnetic second order transition, followed by a
magnetic-magnetic first order transition), but a topological
change is only present in correspondence of the
paramagnetic-magnetic transition. The second phase transition
seems not to have signature in the topology. {\it iii}) A
qualitative agreement between saddle and thermodynamic entropy is
obtained (comparing positive/negative curvature regions), although
the quantitative discrepancy does not allow one to predict the
presence of a first-order transition from topology for certain
values of coupling parameters (taking into account inherent
saddles do not modified the discrepancy, even though a better
quantitative comparison is obtained). Basically, the results of
this paper point toward a weakening of the link between
"thermodynamic" and "topology of the potential energy function"
in mean-field systems. 
A richer scenario is beginning to appear, and a comprehensive
picture is not presently at hand.

The paper is organized as follow. In Sec. II we will introduce the
model in its general form, in Sec. III we will derive the
canonical thermodynamics focusing on two particular cases, in Sec.
IV the topology will be analyzed, calculating stationary point
properties. Conclusions will be drawn in Sec. V.

\section{The model}

We consider a class of mean-field Hamiltonians of the form
\beq
{\cal H} = \sum_{k=1}^{M} H_{2k}
\label{Htot}
\eeq
\noindent with
\beq H_{2k} = - \frac{J_{2k}}{N^{2k-1}} \sum_{\{i\},\{j\}} \cos
(\f_{i_1}+\dots+\f_{i_k} - \f_{j_1}-\dots-\f_{j_k} ) \ ,
\label{Hk}
\eeq
\noindent where the sum is over the sets $\{i\} = i_1,\dots,i_k$
and $\{j\} = j_1,\dots,j_k$ ($i,j=1,\dots,N$) and $\{\f_i\}$ are
angular variables $\f_i \in [0,2\pi)$. We restrict to
ferromagnetic interactions, i.e. $J_{2k}>0$. The system described
by Hamiltonian (\ref{Htot}) can be viewed as an ensemble of $2d$
rotors interacting through mean-field potentials. For $k=1$ the
Hamiltonian (\ref{Hk}) reduces to the usual $XY$ mean-field
Hamiltonian \beq H_{2} = - \frac{J_{2}}{N} \sum_{i,j}
\cos{(\f_{i}- \f_{j})} \ , \label{XY} \eeq while for $k>1$ we have
$2k$-body interaction terms. For example, for $k=2$ 
\beq
H_{4} = - \frac{J_{4}}{N^3}
\sum_{\substack{i_1,i_2\\j_1,j_2}} \cos{(\f_{i_1} + \f_{i_2} -
\f_{j_1}- \f_{j_2})}
\eeq
\noindent which has been recently introduced as a model for
mode-locking laser Hamiltonian \cite{modelock}. Just to give a
further example that will be useful in the following, we
explicitly write also the $k=3$ case
\beq
H_{6} = - \frac{J_{6}}{N^5} \sum_{\substack{i_1,i_2,i_3 \\
j_1,j_2,j_3}} \cos{(\f_{i_1} + \f_{i_2} +\f_{i_3}- \f_{j_1}-
\f_{j_2} -\f_{j_3} )}
\eeq

In this paper we will mainly focus on the case of $2$-terms contributing
to Hamiltonian (\ref{Htot})
\beq
{\cal H} = H_2 + H_{2k} \ ,
\eeq
with $k=2$ and $k=3$, i.e.
${\cal H} = H_2 + H_4$
and
${\cal H} = H_2 + H_6$.

\section{Thermodynamics}
In this section we derive the canonical thermodynamics.\\
Introducing the complex variable $z$:
\beq
z = \rho e^{i \psi} \doteq \frac1N \sum_{i=1}^{N} e^{i \f_i}
\ ,
\label{zdef}
\eeq
\noindent the Hamiltonian (\ref{Htot}) can be written
as
\begin{figure}[t]
\begin{center}
\includegraphics[width=0.43\textwidth]{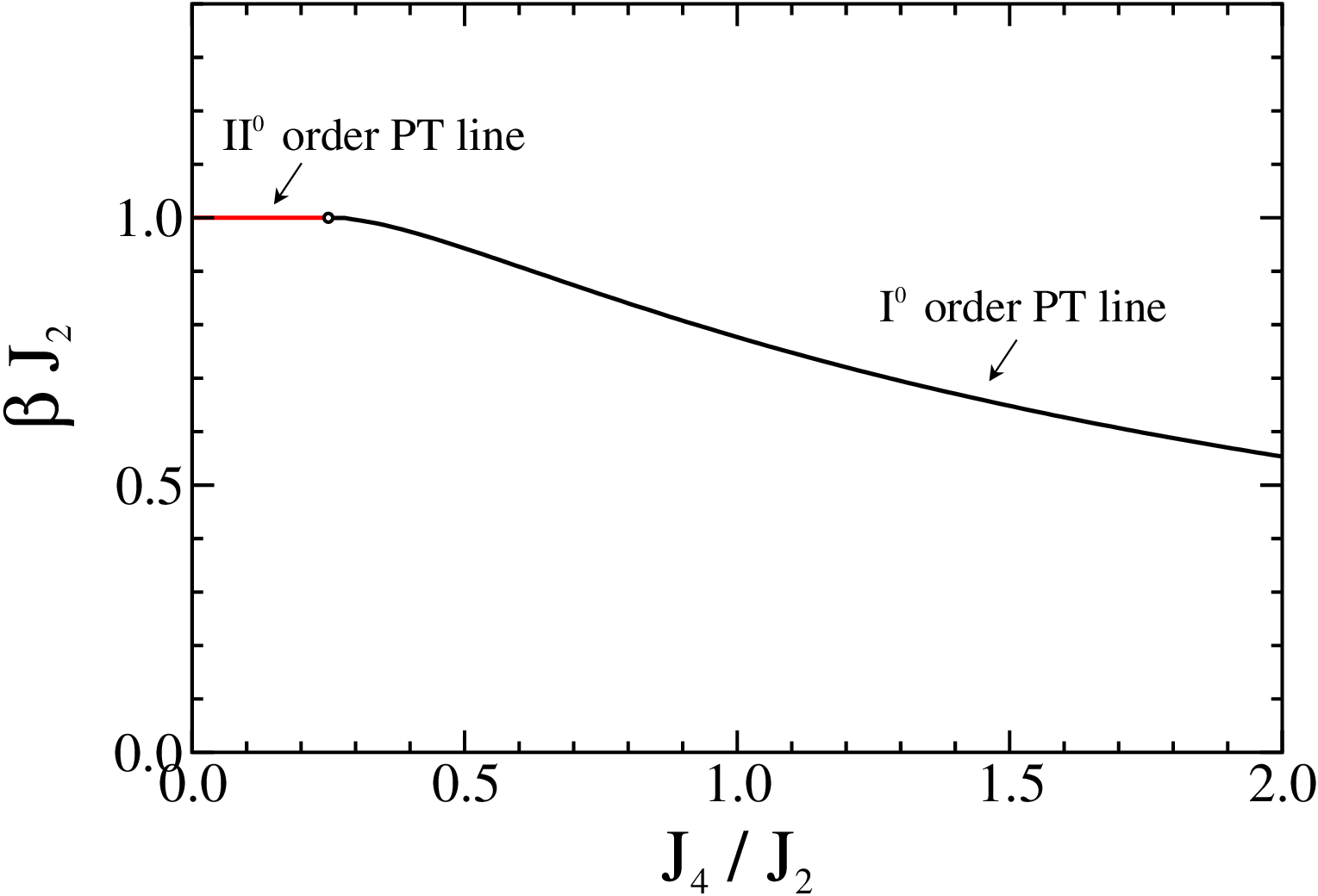}
\end{center}
\caption{(Color online)
$H_2+H_4$ case.
Phase diagram in the ($J_4/J_2,\b J_2$) plane.
Second and first order phase transition (PT) 
lines are reported.}
\label{fig1}
\end{figure}
\beq
{\cal H} = - N \sum_{k=1}^{M} J_{2k}\ \rho^{2k} \ .
\label{Htot2}
\eeq
\noindent It is now possible to write the partition function
\beq
{\cal Z} = \int d\{\f\} \ e^{- \b {\cal H}}  \ ,
\eeq
\noindent as \cite{ktrig_pre}:
\beq \label{partf} {\cal Z} \propto \int d\r \ e^{-N
g(\rho;\beta)} \ , \eeq
\noindent where $\b=1/k_BT$ ($k_B$ is the Boltzmann constant, in
the following we set $k_B=1$) and the function $g(\rho;\beta)$ is
explicitly written as
\beq
 g(\rho;\beta) = \beta \sum_{k=1}^{N} (2k-1) J_{2k} \rho^{2k} - \ln I_o
\left( 2\beta\sum_{k=1}^{M} k J_{2k}\rho^{2k-1} \right) \ ,
\label{g}
\eeq
\noindent where $I_o(x)$ is the modified Bessel function of order
$0$.

Performing the thermodynamic limit $N\to \infty$, the saddle-point
solution dominates the integral in (\ref{partf}). The saddle point
equation is written as:
\beq \r = \frac{I_1(2\b \sum_k k J_{2k} \r^{2k-1})} {I_o(2\b
\sum_k k J_{2k} \r^{2k-1})} \ . 
\label{saddle} \eeq
\noindent This equation can have many solutions for $\r$, that
with the lowest free energy is the stable one
\beq
\beta f_{eq} (\beta) = \min_{\rho} g(\rho;\beta) \ .
\eeq
\begin{figure}[t]
\begin{center}
\includegraphics[width=0.43\textwidth]{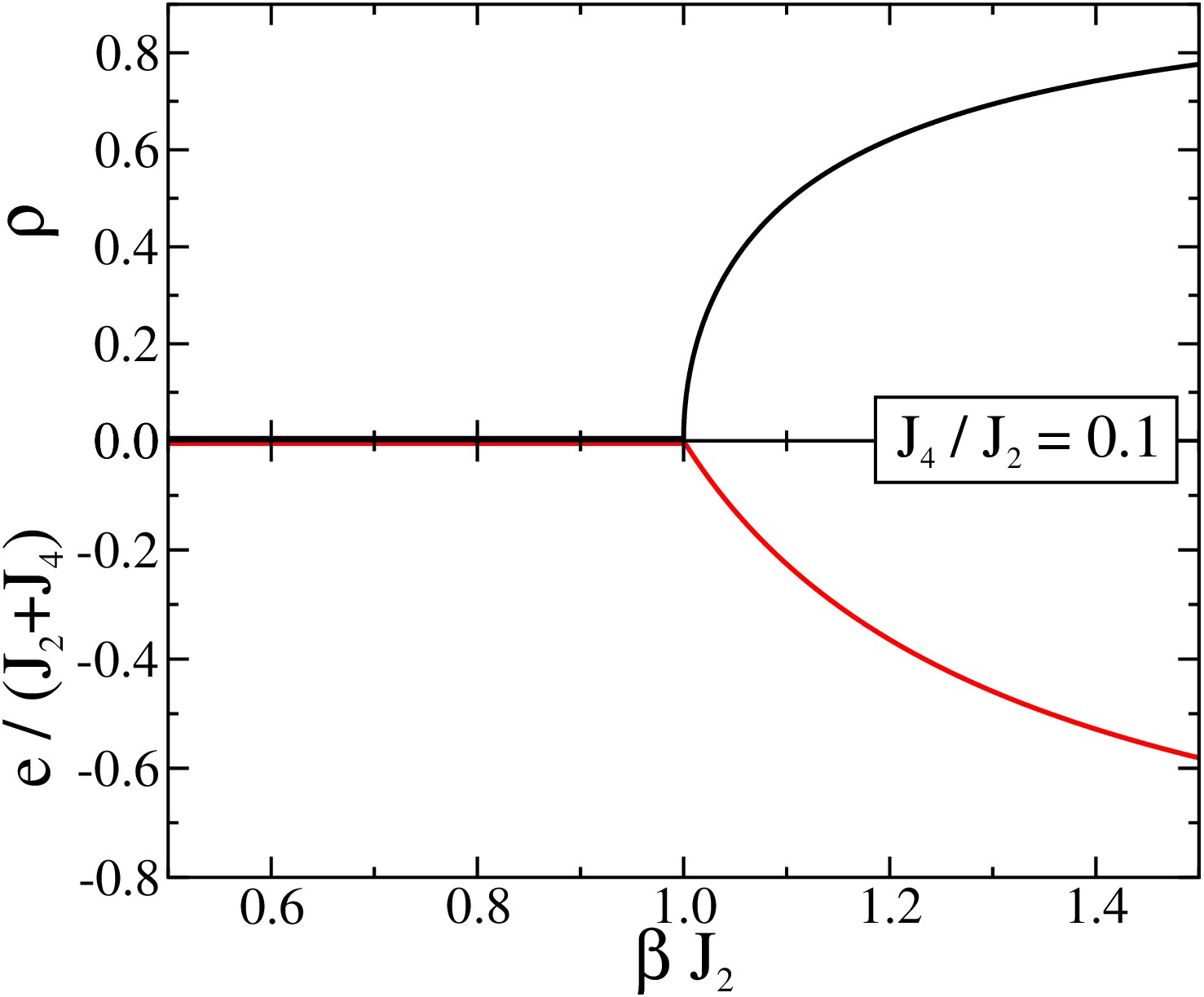}
\end{center}
\caption{(Color online)
$H_2+H_4$ case.
Equilibrium magnetization $\r$ (upper panel)
and energy $e$ (rescaled by $J_2+J_4$, lower panel)
as a function of inverse temperature $\b$ (in unit of $J_2$)
for $J_4/J_2=0.1$. Second-order phase transition takes place at $\b J_2=1$.}
\label{fig2}
\end{figure}
\noindent Thermodynamic properties are obtained from the
$\b$-dependence of $g$. We note that the "paramagnetic" ($\r=0$)
solutions always exist and is stable for small $\b$ value. On
increasing $\b$ solutions with $\r \neq 0$ becomes possible and,
eventually, stable.

We note that for $J_2=0$ only first order phase transitions are
possible. Indeed, in order to have a second order transition the
curvature of the free energy, or of the function $g$, has to
change around the paramagnetic solution $\r =0$, from positive to
negative. For $J_2>0$, close to $\r=0$, we can expand Eq.
(\ref{saddle}) obtaining:
\beq
g \sim \b J_2 (1-\b J_2) \r^2 \ ,
\eeq
\noindent then a second-order phase transition takes place at $\b
J_2 =1$ (if not prevented by a first-order phase transition at
higher temperature). For $J_2=0$ instead, denoting with $J_p$ the
first non-zero term ($p\geq 4$), the expansion of Eq. (\ref{saddle})
reads:
\beq
g \sim \b J_p \r^p \ ,
\eeq
\noindent the curvature is always positive and then second-order
transition cannot take place.

For a single term Hamiltonian ${\cal H}=H_{2k}$, the system
undergoes second- ($k=1$) or first- ($k\geq 2$) order phase
transition. We do not analyze these cases here, as the
thermodynamic/topology relationship falls into the same class of
similar model systems - $XY$ \cite{XY} and $k$-trigonometric
\cite{ktrig} already discussed in the literature. We focus our
attention in many-terms Hamiltonian, taking into account two
interesting cases.

\begin{figure}[t]
\begin{center}
\includegraphics[width=0.43\textwidth]{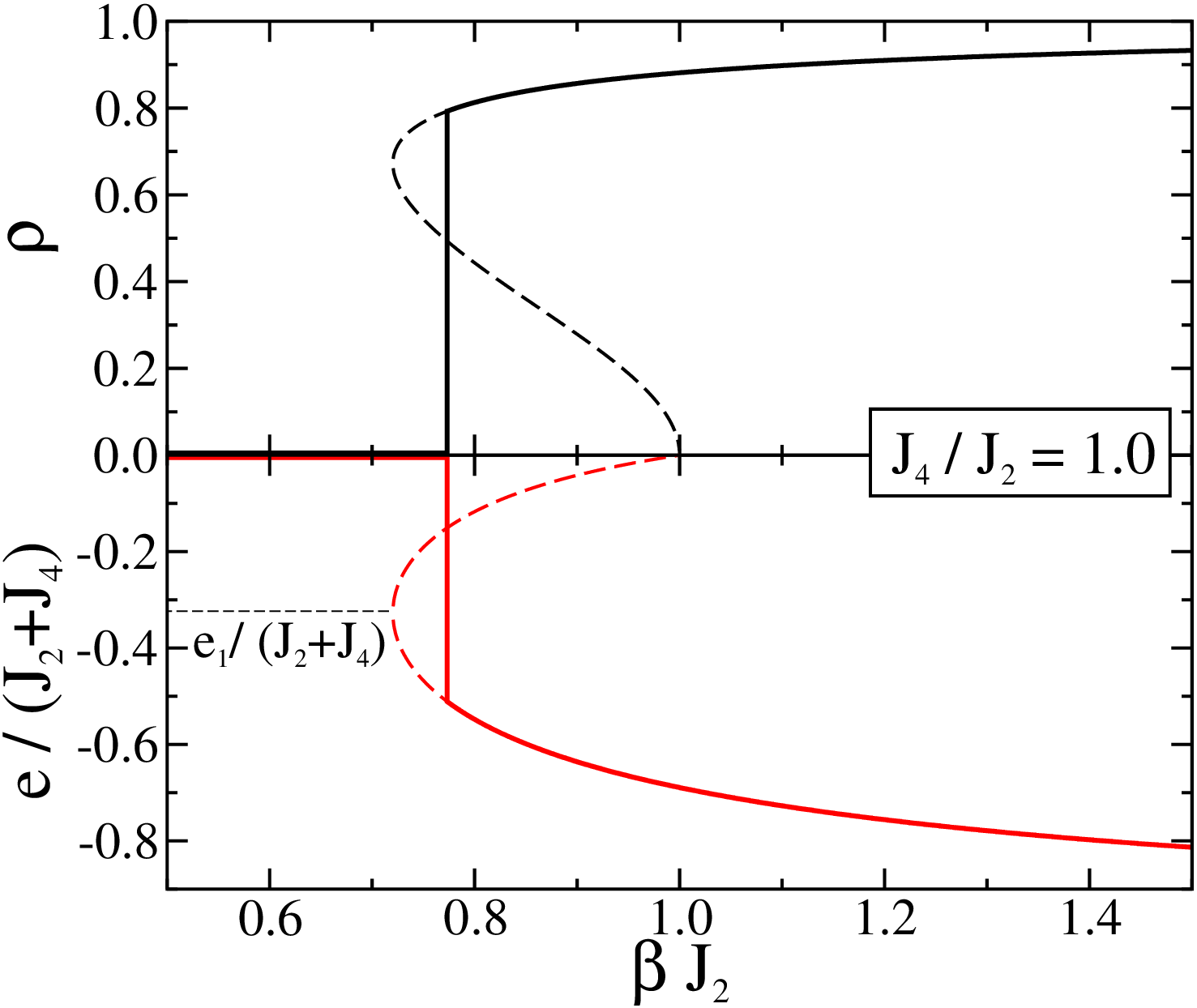}
\end{center}
\caption{(Color online) $H_2+H_4$ case. 
Equilibrium (full lines) and metastable (dashed lines)
magnetization $\r$ (upper panel) and energy $e$ 
(rescaled by $J_2+J_4$, lower panel) as a
function of inverse temperature $\b$ (in unit of $J_2$) for
$J_4/J_2=1.0$. A first-order phase transition takes place at $\b
J_2 \simeq 0.77$. $e_1$ is the energy corresponding to the
appearance of metastable free-energy states. } 
\label{fig3}
\end{figure}

\subsection{$H_2+H_4$ case}
Considering the Hamiltonian ${\cal H}=H_2+H_4$, similarly to the
corresponding Ising case \cite{Ising24}, the plane spanned by the
two coupling parameters is split in two regions (see Fig.~\ref{fig1}), 
which, as can be seen from Eq. \ref{saddle}, are only
determined by the ratio $J_4/J_2$. Specifically,

{\it i)} For $J_4/J_2<1/4$ a second order phase transition takes
place at $\b J_2 = 1$. As an example, in Fig.~\ref{fig2} the
equilibrium magnetization $\rho$ (upper panel) and energy per particle
$e=-J_2\r^2 - J_4 \r^4$ (normalized by $J_2+J_4$, lower panel) are
reported as a function of the (scaled) inverse temperature $\b
J_2$ for the specific case $J_4/J_2=0.1$.

{\it ii)} For $J_4/J_2>1/4$ the transition becomes first order,
with a jump in both magnetization and energy - full lines in
Fig.~\ref{fig3} where the same quantities as in Fig.~\ref{fig2}
are reported for the specific case $J_4/J_2=1.0$. The dashed lines
in Fig.~\ref{fig3} represent metastable states: metastable minimum
of free energy for $e<e_1$ and local maximum for $e>e_1$. 

\noindent For the particular case  $J_4/J_2=1/4$  a tricritical 
point is present at $\beta J_2 =1$.

\begin{figure}[t]
\begin{center}
\includegraphics[width=0.43\textwidth]{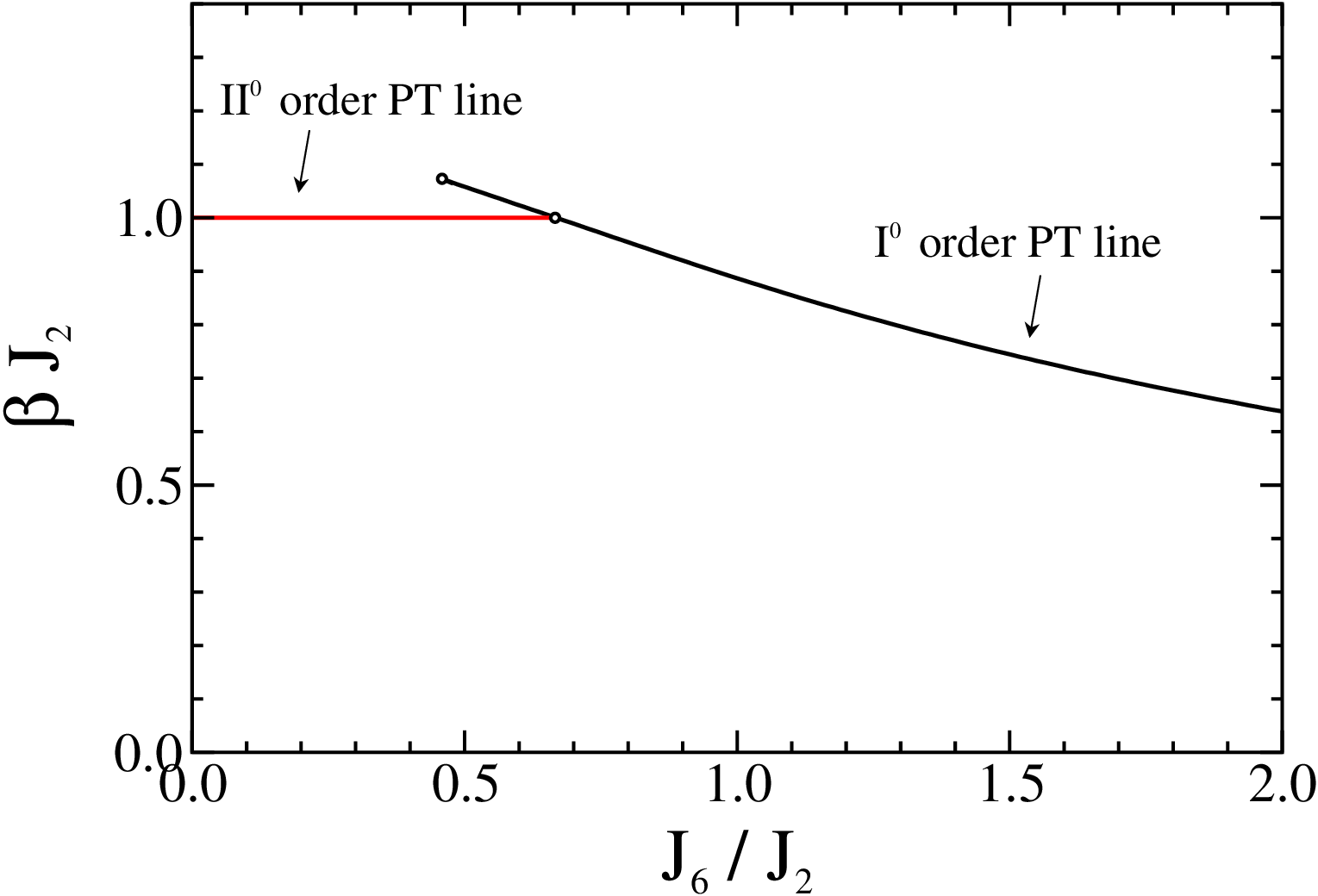}
\end{center}
\caption{(Color online)
$H_2+H_6$ case.
Phase diagram in the ($J_6/J_2,\b J_2$) plane.
Second and first order phase transition (PT)
lines are reported.}
\label{fig4}
\end{figure}

\subsection{$H_2+H_6$ case}
The case ${\cal H}=H_2+H_6$ has a richer phenomenology 
(see Fig.~\ref{fig4}).

\begin{figure}[t]
\begin{center}
\includegraphics[width=0.43\textwidth]{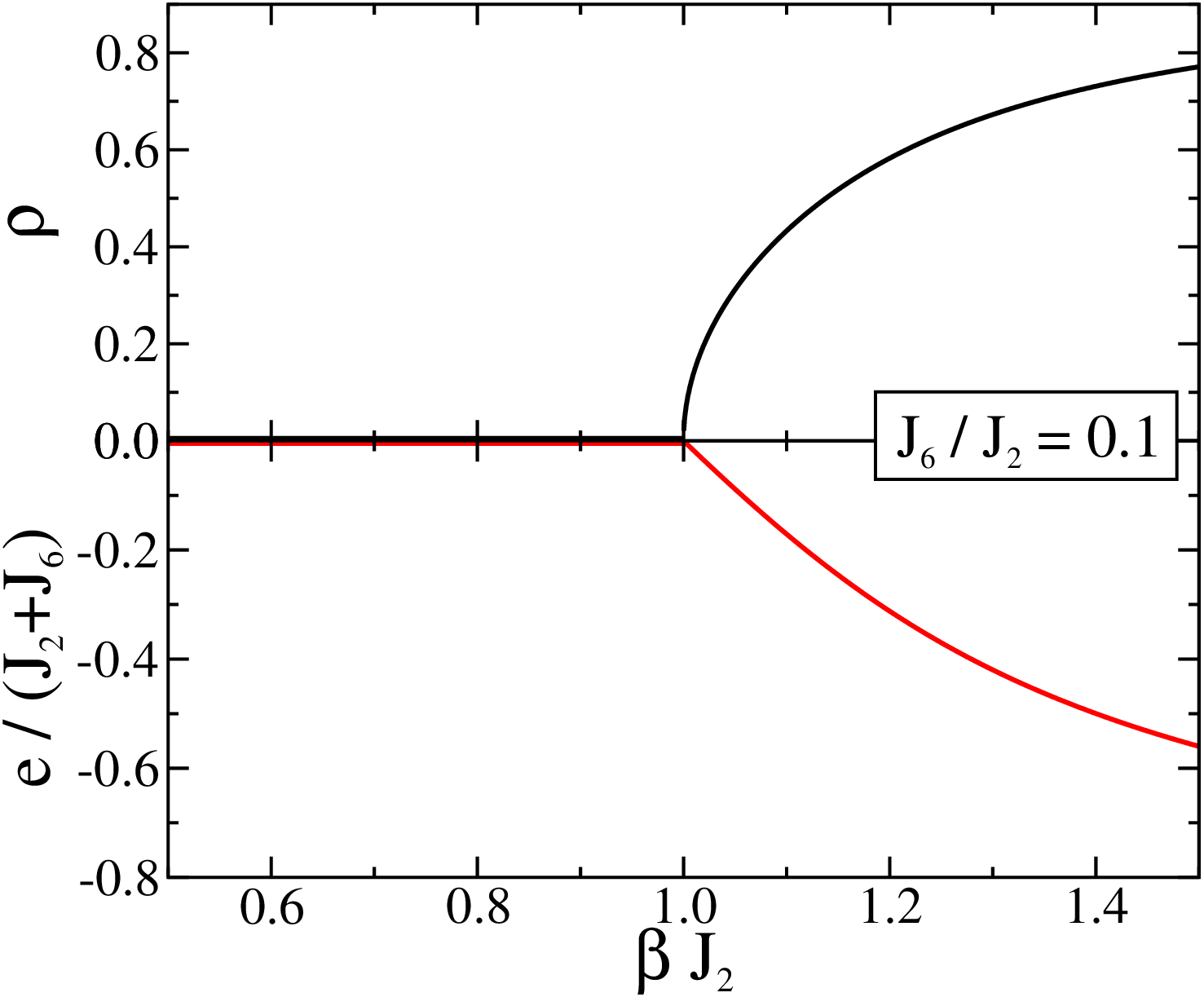}
\end{center}
\caption{(Color online) $H_2+H_6$ case. Equilibrium magnetization
$\r$ (upper panel) and energy $e$ (normalized by $J_2+J_6$, lower
panel) as a function of inverse temperature $\b$ (in unit of
$J_2$) for $J_6/J_2=0.1$. Second-order phase transition takes
place at $\b J_2=1$.} 
\label{fig5}
\end{figure}

{\it i)} For $J_6/J_2< 0.46$ the system, similarly to the case
$H_2+H_4$, undergoes a second order phase transition at $\b J_2=1$.
Fig.~\ref{fig5} reports, as an example, the order parameter and
the energy as a function of the scaled temperature for the
specific case $J_6/J_2=0.1$.

\begin{figure}[t]
\begin{center}
\includegraphics[width=0.43\textwidth]{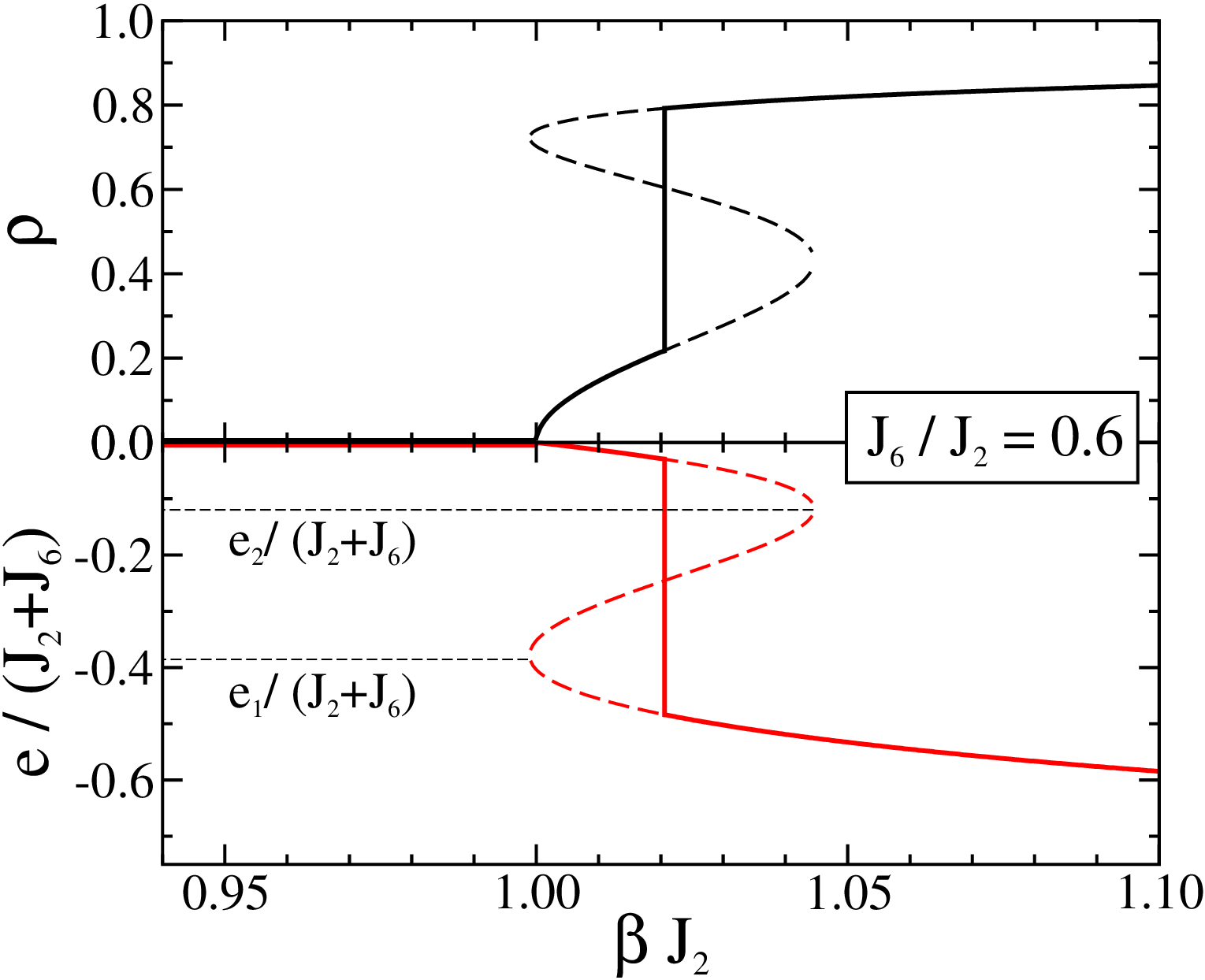}
\end{center}
\caption{(Color online) $H_2+H_6$ case. Equilibrium (full lines)
and metastable (dashed lines) magnetization $\r$ (upper panel) and
energy $e$ (normalized by $J_2+J_6$, lower panel) as a function of
inverse temperature $\b$ (in unit of $J_2$) for the specific case
$J_6/J_2=0.6$. A second-order phase transition takes place at $\b
J_2=1$, followed by a first-order one at $\b J_2 \simeq 1.02$.
$e_1$ and $e_2$ are the energies corresponding to the appearance
of metastable free-energy states. } 
\label{fig6}
\end{figure}

{\it ii)}  For $0.46 < J_6/J_2 < 0.66$ a first order phase
transition occurs after (on increasing $\b$, lowering $T$) 
the second order one
(Fig.~\ref{fig6} reports the specific case $J_6/J_2=0.6$). 

\begin{figure}[t]
\begin{center}
\includegraphics[width=0.43\textwidth]{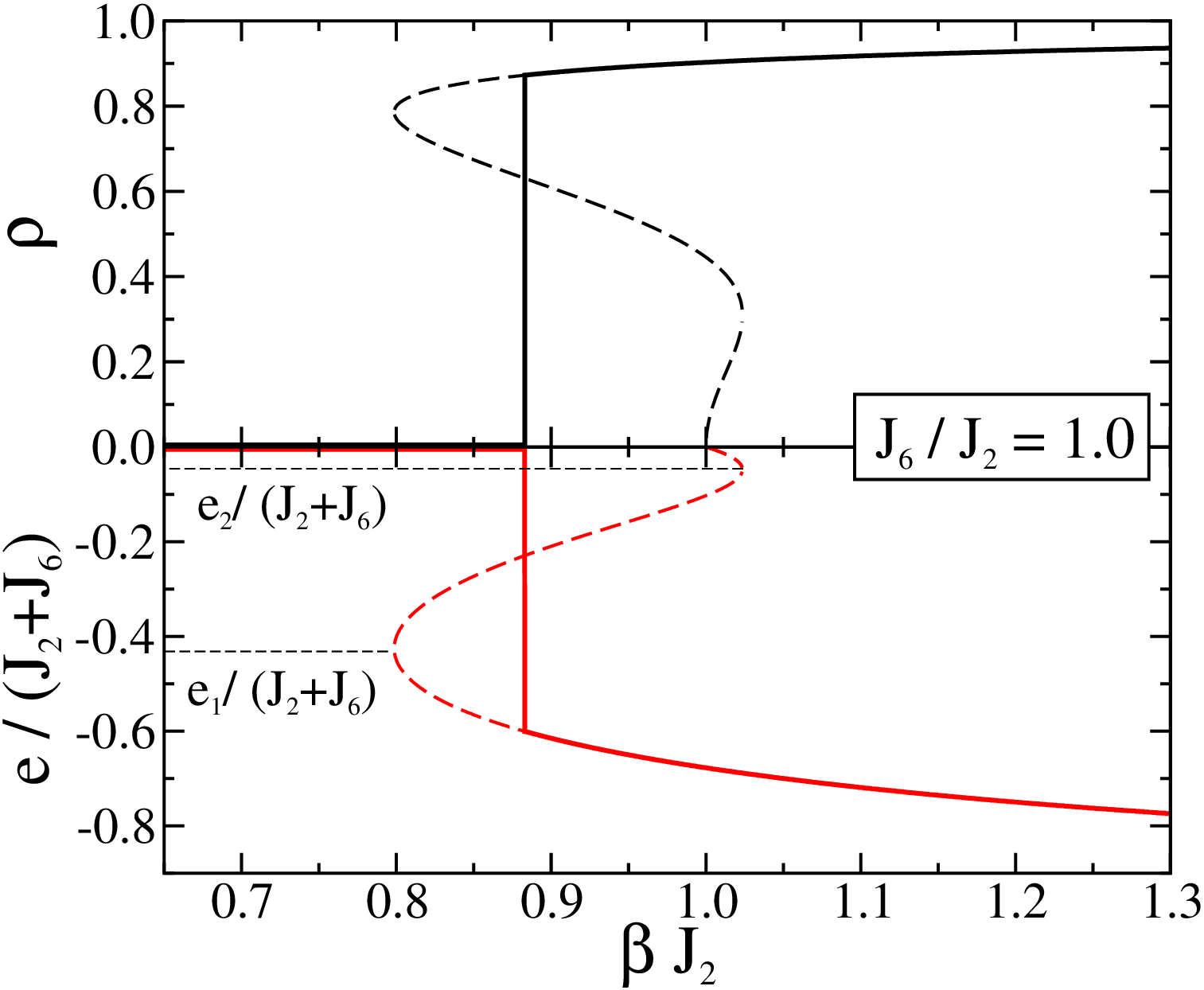}
\end{center}
\caption{(Color online) $H_2+H_6$ case. Equilibrium (full lines)
and metastable (dashed lines) magnetization $\r$ (upper panel) and
energy $e$ (normalized by $J_2+J_6$, lower panel) as a function of
inverse temperature $\b$ (in unit of $J_2$) for the specific case
$J_6/J_2=1.0$. A first-order phase transition takes place at $\b
J_2 \simeq 0.88$. $e_1$ and $e_2$ are the energies corresponding
to the appearance of metastable free-energy states. } 
\label{fig7}
\end{figure}

{\it iii)}  For $J_6/J_2>0.66$ only the first order transition
survives (Fig.~\ref{fig7} reports the specific case
$J_6/J_2=1.0$). 

The dashed lines in Fig.s~\ref{fig6} and \ref{fig7} represent
metastable states: metastable minima for $e<e_1$ and $e>e_2$ and
local maximum for $e_1<e<e_2$.

\section{Topology}

In this Section we will study the topology of the two models
introduced in the previous Section.

We will follow the same line of calculation performed on similar
models in previous works \cite{ktrig,ktrig_long,modelock}. A
quantity directly related to the Euler characteristic of the
manifold $M_e = \{\f | {\cal H}(\f) \le N e\}$, is the
configurational entropy of saddles \cite{ktrig}:
\beq
\s (e)= - n(e) \ln n(e) - (1-n(e)) \ln (1-n(e)) \ ,
\label{sigma_e}
\eeq

\noindent where $n(e)$ is the fractional saddle order (i.e. the
fraction of negative curvatures at the saddle points that are
found in the potential energy hyper-surface when $V(q)=Ne$).
Indeed, we can make the following arguments.\\
A stationary point is defined by 
\begin{equation}
\frac{\partial H}{\partial\varphi_j}=
\sum_{k=1}^{M} 2k J_{2k}\ \rho^{2k-1}
\sin(\varphi_j-\psi)=0 \ \ \ \ \ \   \forall j
\label{staz}
\end{equation}
\noindent The solutions with $\rho>0$ (those with $\rho=0$ are located at $e=0$ energy) are obtained from
$\sin(\varphi_j-\psi)=0$ (for all $j$), then
\begin{equation}
\varphi_j=\left[\psi+ m_j\pi\right]_{{\rm mod}\ 2\pi} 
\label{fij}
\end{equation}
\noindent where $m_j=\{0,1\}$.
Substituting this solution in Eq.~(\ref{zdef}) we obtain
\begin{equation}
\label{rho_n}
\rho=1-2n  \ ,
\end{equation}

\noindent where 
\begin{equation}
n=\frac1N \sum_j m_j
\label{fso}
\end{equation}
is the fractional saddle order (as it will be clear soon)
and we have used the identity $(-1)^{m_j}=1-2m_j$.
We conclude that there are no stationary points with $n>1/2$ ($\rho$ is positive defined).
The order of a stationary point is defined by its downward curvatures, i.e. by the number 
of negative eigenvalues of the Hessian matrix 
$H_{ij}=\partial{^2H} / \partial{\varphi_i}\partial{\varphi_j}$.
It is possible to show that in the thermodynamic limit the Hessian becomes diagonal
\begin{equation}
H_{ij} \simeq \delta_{ij}\ \lambda_{j} \ ,
\label{Hess}
\end{equation}
\noindent where
\begin{equation}
\begin{split}
\lambda_{j} &= \cos(\varphi_j-\psi)\  \sum_{k=1}^{M} 2k J_{2k}\ \rho^{2k-1} \\
&= (-1)^{m_j}\  \sum_{k=1}^{M} 2k J_{2k}\ \rho^{2k-1} \ .
\end{split}
\end{equation}
\noindent Therefore, the saddle order is given by the number of $m_j=1$ at the considered 
saddle point, and the fractional saddle order $n$ is then given by Eq. (\ref{fso}).
Moreover, the number of saddles with a given $n$ is given by the binomial coefficient,
then
\begin{equation}
\sigma(e)=\lim_{N\to\infty}\frac{1}{N}\ln\binom{N}{Nn(e)} \ ,
\end{equation}
\noindent from which the Eq.~(\ref{sigma_e}) follows straightforward.
The latter can be written in the form
\beq
\s (e)= - \frac{1-\rho(e)}{2} \ln \frac{1-\rho(e)}{2} -
\frac{1+\rho(e)}{2} \ln \frac{1+\rho(e)}{2}\
\eeq
\noindent where $\r(e)$ is obtained from the thermodynamics, i.e.
from the solution of the equation:
\beq \label{exi} e = - \sum_{k=1}^{M} J_{2k}\ \rho^{2k}(e)  \eeq

\noindent We note that the quantity $\s$ is singular at
$e=0$, due to the fact that $e>0$ is a forbidden energy region,
so $\sigma$ has a discontinuity, jumping from 
a finite value to zero.

As discussed in the introduction, a central quantity in the
comparison between thermodynamic and topology is the curvature of
$\s(e)$. Specifically, we are interested in finding the energy
values where there is changes in curvature of $\s(e)$, i.e. the
energies where the second derivative of $\s(e)$ vanishes.
\noindent After some algebra we get
\beq \frac{d^2\s}{de^2} = - \frac12 \left( \frac{d\rho}{de}
\right)^2 s[\rho(e)] \ , \label{d2s} 
\eeq
\noindent where the function $s$, which depends on $e$ through
$\r(e)$, is:
\beq
s(\rho)= \frac{1}{1-\rho} + \frac{1}{1+\rho} + \frac{\sum_k
2k (2k-1) J_{2k} \rho^{2k-2}}{\sum_k 2k  J_{2k} \rho^{2k-1}} \ln
\frac{1-\rho}{1+\rho}
\eeq
\noindent From Eq.~\ref{d2s} we have
\beq s \lesseqgtr 0 \; \Longleftrightarrow \; \frac{d^2\s}{de^2}
\gtreqless 0 \ , \eeq
\noindent that is upward, null, downward curvature respectively.
Studying the positivity of $s$ allows us to determine the curvature
of $\s$. We now specialize the calculations to the two previous
cases.

\begin{figure}[t]
\begin{center}
\includegraphics[width=0.45\textwidth]{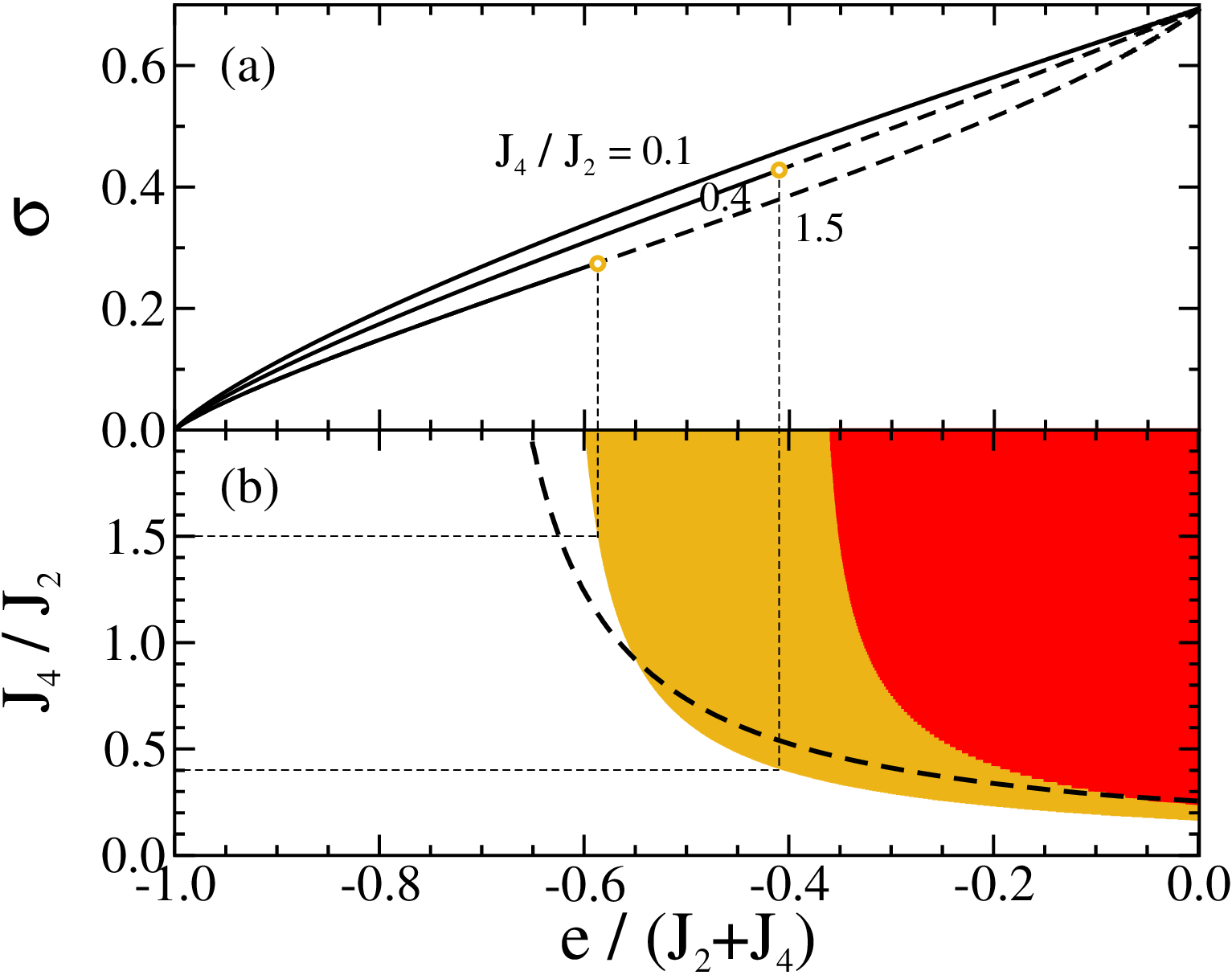}
\end{center}
\caption{(Color online) $H_2+H_4$ case. (a) Saddle-entropy $\s$ as
a function of energy $e$ (normalized by $J_2+J_4$) for
$J_4/J_2=0.1, 0.4, 1.5$. Full (dashed) lines correspond to
negative (positive) curvature. Symbols mark turning points. (b)
Couplings-energy plane: light-grey region corresponds to $\s$
positive curvature, dark-grey region to entropy $S$ positive
curvature. Dashed line is the inherent saddle counterpart of the
border of dark region.} \label{fig8}
\end{figure}

\subsection{$H_2+H_4$ case}
In Fig.~\ref{fig8}, upper panel, the quantity $\s(e)$ is plotted
as a function of energy $e$ (normalized by $J_2+J_4$) for three
selected case $J_4/J_2=0.1, 0.4, 1.5$. Full lines correspond to
negative curvature, while dashed lines to positive (symbols
represent turning points). In the $J_4/J_2=0.1$ case 
(second order phase transition located at $e=0$) 
the curvature is always negative, while
in the other two both regions are present. The quantity $\s(e)$
is singular (discontinuous) at $e=0$,
corresponding to the
thermodynamic transition energy.
In Fig.~\ref{fig8}, lower panel, the plane $(J_4/J_2,e)$ is drawn
(the energy is normalized by $J_2+J_4$). The light-grey region
corresponds to positive curvature of $\s(e)$, its border being the
null curvature line - the correspondence with the turning points
of $\s(e)$ in the upper panel of Fig.~\ref{fig8} is evidenced with
the thin dashed lines for $J_4/J_2=0.4$ and $1.5$. The border of
the light-grey region does not intersect the value $J_4/J_2=0.1$,
and, therefore, no turning point exists for such a $J_4/J_2$
value.

The dark-grey region (that fully lies inside the light-grey one)
represents the thermodynamically forbidden region $e>e_1$ (see
Fig.~\ref{fig3}). It corresponds to positive curvature of
thermodynamic entropy $S(e)$. If the topological hypothesis was
correct, the two, light- and dark-grey, regions would coincide.
This is not the case, suggesting a not one-to-one correspondence
between thermodynamic and saddle entropy.

In the study of other models \cite{1d,phi4}, when the energy of the
topological change were not coincident with the energy of the
phase transition, it was found that the ``weak'' topological
hypothesis was correct. Indeed, it has been shown that the
correspondence between topological change and phase transition
energies were obtained considering inherent saddle properties: the
energy of topology transition has been found to correspond to the
inherent saddle energy. The latter quantity was obtained
minimizing the quantity $W=|\nabla V|^2$ \cite{1d,ktrig_long,sad_prl}.

\begin{figure}[t]
\begin{center}
\includegraphics[width=0.45\textwidth]{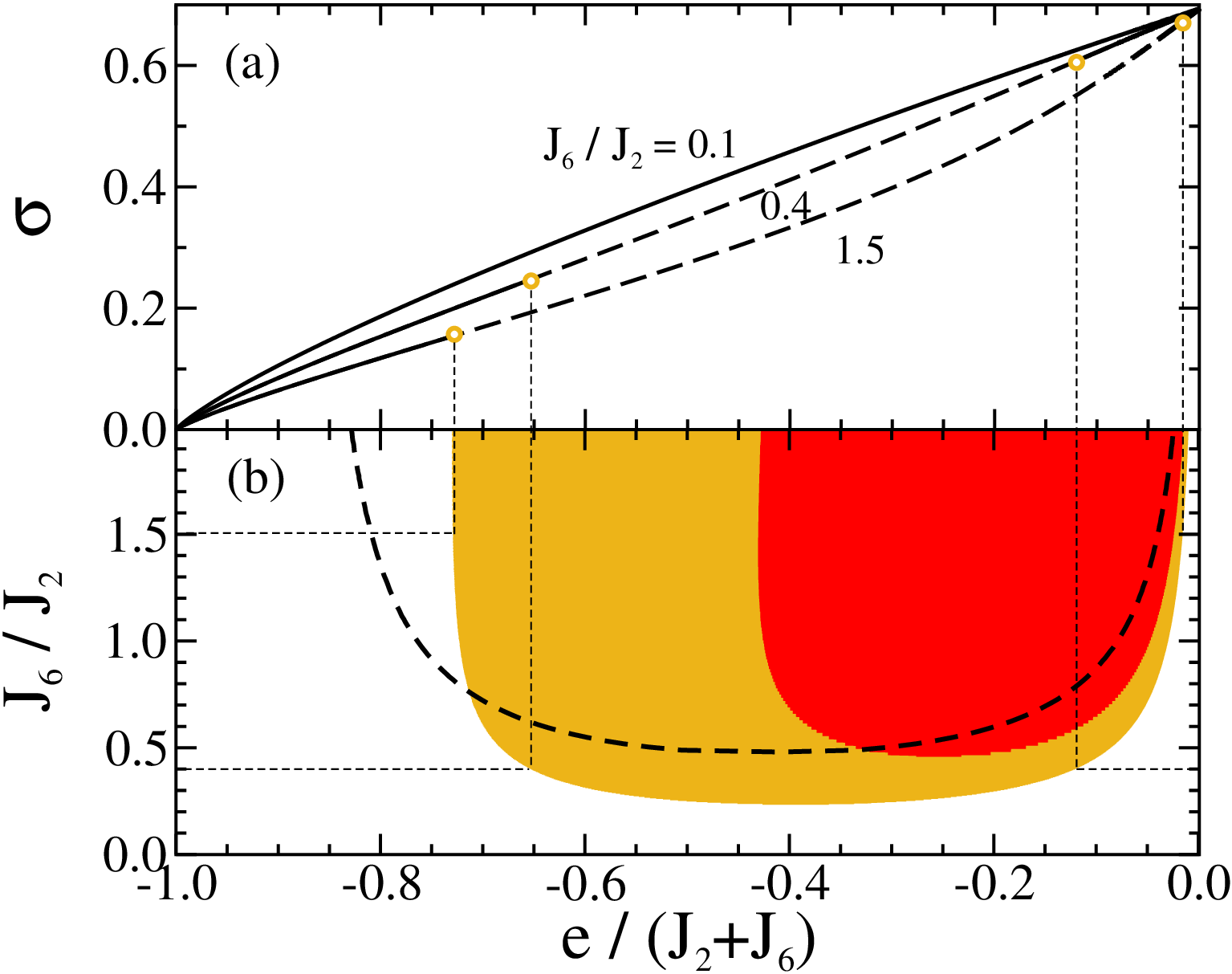}
\end{center}
\caption{(Color online) $H_2+H_6$ case. (a) Saddle-entropy $\s$ as
a function of energy $e$ (normalized by $J_2+J_6$) for the
specific cases $J_6/J_2=0.1, 0.4, 1.5$. Full (dashed) lines
correspond to negative (positive) curvature. Symbols mark turning
points. (b) Couplings-energy plane: light-grey region corresponds
to $\s$ positive curvature, dark-grey region to entropy $S$
positive curvature. Dashed line is the inherent saddle counterpart
of the border of the dark-grey region.} \label{fig9}
\end{figure}

Dashed line in Fig.~\ref{fig8}b is the inherent saddle line
obtained from the border of the dark-grey region (zero curvature
thermodynamic entropy) applying the minimization procedure of $W$.
The latter has been performed solving steepest descent equations
of the form $\dot{\varphi} = - \nabla W$ starting from 
equilibrium initial configurations and taking at the end the infinite
time limit. Following similar calculations of Ref. \cite{ktrig_long},
we obtain the saddle energy $e_s=e(\rho_\infty)$ where 
$
\rho_\infty = 
L_0(- \beta de/d\rho ) / 
I_0(- \beta de/d\rho )
$
and $L_0$ is the
modified Struve function of order $0$:
$L_0 (\alpha) = 2 \pi^{-1} \int_0^{\pi/2}  d\varphi \sinh (\alpha \cos \varphi)$.
A better correspondence is
obtained between the light-grey border and the dashed line,
however the two regions are not yet coincident. This indicates that
the "weak" topological hypothesis could only be considered a good
approximation, but not a quantitative prescription for the
location of the phase transitions.
A comment is in order.
Although obtained from canonical ensemble, the saddle energy map
(from instantaneous to saddle energy) is expected to be 
ensemble-independent, when extrapolated to thermodinamically
forbidden energy regions. 
``Ensemble inequivalence'' phenomena \cite{eip} are then
supposed to not affect the results.

Summarizing: the singular behavior of $\s$ at $e=0$ signals the
presence of a thermodynamic transition, the curvature of $\s$ is
``quasi''-related to the presence of thermodynamically forbidden
region and then to the appearance of a first-order transition. The
``quasi''-relation is due to the fact that there are regions where
the curvature of $\s$ is positive and the curvature of
thermodynamic entropy is negative.

\subsection{$H_2+H_6$ case}
In Fig.~\ref{fig9} the same as in the previous case is reported
for ${\cal H}=H_2+H_6$. In the upper panel the energy dependence
of $\s$ is reported for the specific cases $J_6/J_2=0.1, 0.4,
1.5$. As before, dashed lines correspond to positive curvature
regions. In Fig.~\ref{fig9}b the plane $(J_6/J_2,e)$ is drawn:
light-grey region is the $\s$-positive curvature, dark-grey one is
the $S$-positive curvature. Dashed line is the inherent saddle
counterpart of dark region border. Again, beside an overall
qualitative behavior, we do not find a quantitative correspondence
between the two regions that, according to the weak topological
hypothesis, should coincide. We note that there are values of
parameters ($0.2\lesssim J_6/J_2 \lesssim 0.5$) where the
curvature of $\s$ is upward (for certain energy values) and a
second order transition takes place (the curvature of $S$ is
downward). 

\section{Conclusions}

By analytically studying the thermodynamics and the topology of
mean-field models obtained by the sum of two different interaction
terms in the Hamiltonian (\ref{Htot}) - $2+4$ and $2+6$ body terms
-  we are able to compare thermodynamic and topological quantities and
test the validity of both the ``topological hypothesis''
\cite{cccp} and the ``weak topological hypothesis'' \cite{1d}. The
models have a rich phase space structure. The $H_2+H_4$ model
performs second or first order phase transition depending on the
coupling parameters values (a tricritical point joins the two).
The $H_2+H_6$ model has a further possible behavior: it can
undergo a double phase transition, a second order followed by a
first order one. Topological invariant (saddle entropy) has a
change in correspondence of the zero energy (the paramagnetic
energy), so confirming in that case the equivalence between
topological and thermodynamic transition points. 
However, only one topological change occurs, also for 
the $H_2+H_6$ model exhibiting two phase transition points.
Indeed, in this case the first order phase transition does not have a
topological counterpart, being the topological saddle entropy
a smooth function at the corresponding phase transition energy values.
It is worth noting that this is a quite unexpected result,
all the models analyzed so far presenting a topological
change at some energy value (even though not coincident
with the phase transition one for some model system).
Future studies should establish if this is a 
mean-field ``pathology'' or has a deeper origin.
Moreover, the curvature of the saddle entropy (as
a function of energy), seems not to be strictly related
to the presence of first order phase transition (as previously
observed in a different model \cite{ktrig}): there are regions in
the parameters values for which positive curvature corresponds to
second order phase transitions. In other words, the curvature
properties of saddle entropy do not coincide with those of the
thermodynamic entropy. Taking into account the possibility that
the relevant energy levels are given by underlying saddles, a
better agreement between curvature regions has been found,
although there are no quantitative coincidence.

In conclusion, for the analyzed mean-field models a topology change is present 
at the same energy level at which a phase transition takes place,
in agreement with the ``topological hypothesis''.
However, the
information encoded in the topology seems not to be sufficient to
predict all the possible thermodynamic behaviors of the system,
like the presence of two phase transition points or the phase
transition order.


\end{document}